# Domain Nucleation and Annihilation in Uniformly Magnetized State under Current Pulses in Narrow Ferromagnetic Wires

Yoshihiko Togawa[1]*, Takashi Kimura[1,2], Ken Harada[1,3], Tetsuya Akashi[3,4], Tsuyoshi Matsuda[1,3], Akira Tonomura[1,3] and Yoshichika Otani[1,2]

[1]*Frontier Research System, the Institute of Physical and Chemical Research (RIKEN), Hatoyam, Saitama, 350-0395, Japan /Hirosawa, Wako, Saitama 351-0198, Japan*
[2]*Institute for Solid State Physics, The University of Tokyo, Kashiwanoha, Kashiwa, Chiba 277-8581, Japan*
[3]*Advanced Research Laboratory, Hitachi, Ltd., Akanuma, Hatoyama, Saitama 350-0395, Japan*
[4]*Hitachi Instruments Service Co., Ltd., Yotsuya, Shinjuku-ku, Tokyo 160-0004, Japan*

**We investigate the current-driven magnetization dynamics in narrow Permalloy wires by means of Lorentz microscopy and electron holography. Current pulses are found to transform the magnetic structure in the uniformly magnetized state below the Curie temperature. A variety of magnetic states including reversed magnetic domains are randomly obtained in low probability. The dynamics of vortices found in most of observed magnetic states seems to play a key role in triggering the magnetization reversal.**

*Corresponding author. E-mail address: yoshihiko.togawa.tq@hitachi.com

Manipulation of magnetic domain structures in ferromagnetic submicron wires using spin-polarized current has been a hot subject because of a wide range of possible applications in hard disk drives, magnetic random access memory, and other spintronic devices.[1] Transfer of the spin angular momentum from conduction electrons to the constituent localized magnetic moments induces the spin torque that drives the magnetic domain wall along the electron flow.[2-6] In addition, the spin torque excites the spin-wave fluctuation,[7-9] perturbing the uniformly magnetized state to nucleate a domain wall.[10] This idea will be the use of the spintronic magnetization control.

Experimental understanding of the underlying physics of the current-excited magnetization dynamics relies mainly on macroscopic magneto-resistance measurements and/or imaging techniques that provide image snapshots of static domain walls.[11-16] Although the results of these experiments match up to predictions based on the spin transfer,[2-6] rather complicated dynamics of the domain wall has been directly found in recent imaging studies.[17,18] The local structural deformation of the domain wall impedes its propagation.[17] The domain walls are displaced both along and against the electron flow.[18] Oscillatory domain wall motion is also demonstrated by nanosecond-long current pulses.[19] These observations exhibit a variety of the magnetization dynamics under the spin torque.

In this study, we present microscopic behavior of the magnetization subjected to the pulsed current above the threshold current density $J_{th}$ examined by means of Lorentz microscopy and electron holography. Current pulses are found to nucleate and annihilate chained vortices, crosstie walls and reversed magnetic domains in the uniform state below the Curie temperature $T_C$.[20] A variety of magnetic structures are obtained in low probability, suggesting stochastic nature of the phenomena. This characteristic may be related to an excitation process of the spin-wave and/or thermal fluctuation.

Permalloy zigzag wires with thickness of 30 nm and width of 500 nm are fabricated on 30-nm-thick $Si_3N_4$ membrane for transmission electron microscope (TEM) observation by lift-off technique using electron-beam lithography.[18] Five wires are connected to large Permalloy pads in parallel as shown in Fig.1(a). One of pads is electrically grounded and the positive current pulses with variable magnitude and pulse duration less than 1 μs are applied. Resistivity of Permalloy is 14.6 μΩ cm at room temperature. The magnetization dynamics is investigated by means of Lorentz microscopy and electron holography using the 300 kV field-emission TEM. Lorentz microscopy enables a real-time observation using video that saves enormous time to obtain a sequence of snapshots of the magnetization before and after the current-pulse application. While TEM observation, the wire resistance is simultaneously measured to monitor the wire temperature.[21] The sample is set above the objective lens to avoid the magnetic field produced by the electromagnetic lenses.

The wire resistance as a function of the current density at the pulse duration of 300 ns is shown in Fig.1(b). The kink found at the beginning of a linear dependence above the current density of $2.07 \times 10^{11}$ A/m$^2$ corresponds to $T_C$.[22] Characteristic structural changes in the magnetic states are examined with increasing the current density below $T_C$ as reported before.[18] The first deformation, displacement and annihilation of domain walls are observed in the one of five wires at $1.75 \times 10^{11}$, $1.91 \times 10^{11}$ A/m$^2$ (defined as $J_{th}$), and $1.98 \times 10^{11}$ A/m$^2$, respectively.

Displacement of adjacent domain walls above $J_{th}$ brings about their annihilation. Figure2 shows part of a series of Lorentz micrographs taken each time the current pulse of $2.00 \times 10^{11}$ A/m$^2$ is applied. Two domain walls seen in Fig.2(a) approach, merge and form the magnetic structure consisting of chained vortices as shown in Fig.2(b). Application of subsequent three current pulses transforms the above magnetic structure into different states twice and returns back to the uniformly magnetized state [Fig.2(c)].

Interestingly, repetitive application of current pulse with the same magnitude sometimes induces a drastic structural change even in the uniformly magnetized state.[20] This type of event is observed 58 times while 2000 current pulses are applied. Chained vortices, crosstie and their complex structure [e.g., Fig.2(d)] are observed 7, 35, and 5 times out of 2000 pulses, respectively. More importantly, the magnetization reversal, i.e., the domain nucleation in the uniformly magnetized state is observed 11 times as shown in Fig.2(f). The induced magnetic state immediately vanishes 51 times out of 58 times by subsequent current pulse application [e.g., Figs.2(d)-2(e)]. The transformation of the magnetic state occurs only when the current pulse is applied. Remarkable is that a variety of magnetic structures are produced by current pulses. For instance, while chained vortices and crosstie structure are almost always formed at the same positions, their detailed structures are different as in Figs.2(a) and 2(f) or 2(b) and 2(d). This is in contrast with the observation that the demagnetization process in the applied in-plane magnetic field perpendicular to the averaged direction of the wire always generates the same magnetized structure (transverse domain wall) at every corner of the wire.[18]

Figures 3(a)-3(c) show detailed distribution of magnetic flux lines obtained by electron holography[23] in the same region corresponding to a part of that in Lorentz micrographs in Figs.2(d)-2(f), respectively. There are three representative magnetic domain structures. Figure 3(a) shows that magnetic flux lines run through the induced chained vortices to connect adjacent regions magnetized uniformly in the same direction, while an opposite component of magnetic flux lines are taken into the vortices. This complex magnetic structure is essentially different from the uniformly magnetized domain in Fig. 3(b) and the reversed magnetic domain structure where magnetic flux lines are disconnected by the adjacent domain walls in Fig. 3(c).

The uniformly magnetized state in Fig.3(b) is more stable than other states in Figs. 3(a) and 3(c) since the state in Fig. 3(b) has no flux leakage from the side edge. Energy expense is required for the system to leak out magnetic flux lines to generate vortices and reversed magnetic domains. Thus, the probability of the nucleation may be as small as 3 % for vortices and 0.5 % for domains. For similar energetical reasons, the induced magnetic state vanishes and relaxes into the more stable uniformly magnetized state by a subsequent current pulse in probability of 90 %. This result suggests that a magnetization dynamics is predominantly involved in erasing the induced magnetized structures. Indeed, vortices and antivortices embedded in a crosstie structure prefer to annihilate pairwise.[24] However, this process leaves one vortex at least to preserve the total winding number.[25] Thus, the remaining vortices should be removed to reverse the magnetization. Taking into consideration that vortices alternately stand aside from the center of the wire so that magnetic flux lines meander through the induced magnetic structure in the static state [e.g., Fig. 3(a)], we expect that the uniformly magnetized state is revived when vortices are expelled from the closest wire edges. It should be noted that such lateral movement of vortices can be driven by the spin torque.[26, 27]

Vortices and crosstie structure are frequently nucleated around the wire corner and the notch (a structural defect) in a straight segment of the wire (Fig. 2). The magnetization is naturally bent around such regions even in the uniformly magnetized state as shown in Fig. 3(b), suggesting that the geometry with the spatial gradient of the magnetization favors the domain nucleation. Introduction of deep notch or sharp bend in the wire may thus be useful to raise the probability of the domain nucleation of about 0.5 % in the present geometry. It is also noted that the spin torque is enhanced around such structures, leading to the structural transformation in the uniformly magnetized state with an assistance of the spin-wave fluctuation.[7-10] The process based on the spin-wave fluctuation has stochastic nature, which may be manifest in observed features of the domain nucleation in low probability. Thermal fluctuation is also enhanced due to temperature increase during current pulse application and may trigger stochastic phenomena. Theoretical and numerical analysis of vortex dynamics incorporating both effects of spin-wave and thermal fluctuation will be needed to reveal an exact mechanism of the observed phenomena.

In conclusion, we present detailed behavior of the domain nucleation and annihilation in the uniformly magnetized state induced by current pulses below $T_C$. We hope that present experimental findings will stimulate physical understanding of the current-driven magnetization dynamics and enrich the spintronic technology of the magnetization manipulation.

We thank Dr. J. Shibata of RIKEN for fruitful discussions, and Mr. N. Moriya of Hitachi Ltd. and Mr. T. Furutsu of Hitachi High-Technologies Co. for technical support.

**Figure captions**

Figure 1 (a) TEM image of five Permalloy zigzag wires connected to Permalloy pads. Dashed square on the wire represents the field of view in Fig.2. (b) The wire resistance as a function of the current density flowing in the wire at 300 ns current pulse. Broken line on the data is a guide for the eye. Characteristic changes in the magnetized states, the threshold current density $J_{th}$ and the Currie temperature $T_C$ are indicated by arrows. Closed circles point the current density used in Figs.2 and 3.

Figure 2 Lorentz micrographs showing magnetization changes induced by the application of 300 ns current pulses of $2.00 \times 10^{11}$ A/m$^2$. The electron flows from the right to the left. (a) First current pulse application. (b) Third current pulse application. Chain vortices with crosstie structure are formed from adjacent vortices domain walls in (a). (c) Sixth current pulse erases the induced magnetized state and recovers the uniformly magnetized state. (d) Tenth current pulse nucleates chained vortices with crosstie structure again. (e) Eleventh current pulse revives the uniformly magnetized state. (f) 1280th current pulse reverses the magnetization completely in a part of the wire. The next current pulse erases the nucleated domain and revives the uniformly magnetized state. The direction of the magnetization is indicated by red arrows. Scale bar corresponds to 2 μm.

Figure 3 Twice phase-amplified interference images showing the magnetic flux lines distribution in the same region of the wire induced by the application of 300 ns current pulses of $2.00 \times 10^{11}$ A/m$^2$. The magnetized state in (a)--(c) corresponds to a part of that in Lorentz micrographs in Figs.2(d)-2(f), respectively. The direction of the magnetization is indicated by red arrows. The magnetization is slightly bent around the notch in (b).

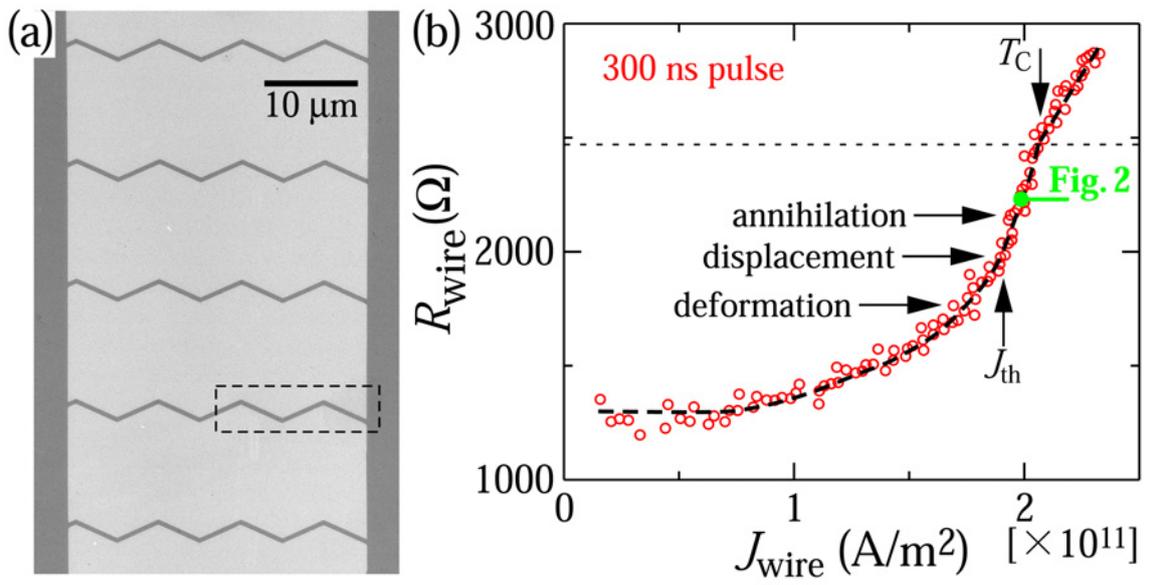

Fig. 1. Y. Togawa *et al*., "Domain Nucleation and Annihilation …"

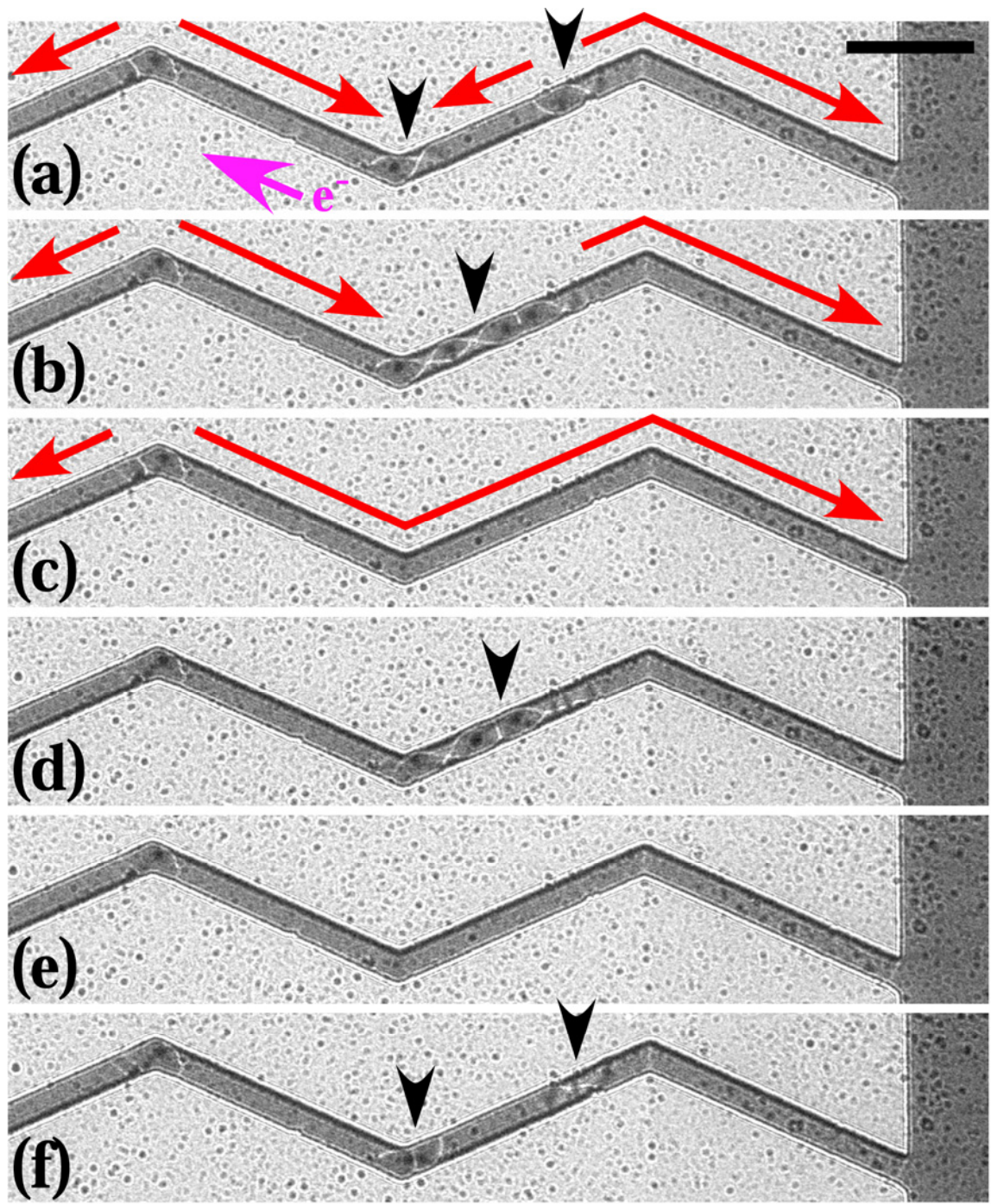

Fig. 2. Y. Togawa *et al.*, "Domain Nucleation and Annihilation …"

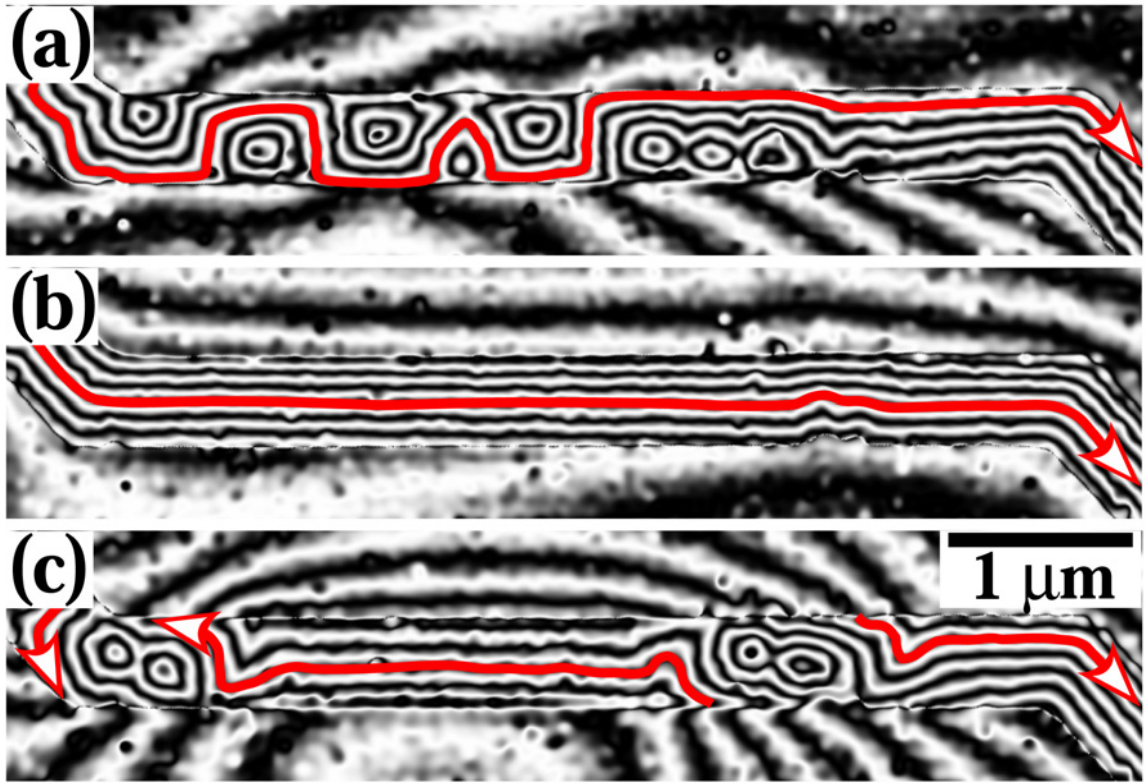

Fig. 3. Y. Togawa *et al*., "Domain Nucleation and Annihilation …"